\documentclass{aa}
\usepackage{natbib}
\usepackage{amssymb}
\usepackage{amsmath}
\usepackage{graphicx}
\def\revised#1{#1}
\def\aap{A\& A}

\def\nat{Nature}
\def\apj{ApJ}

\def\apjl{ApJL}
\def\apjs{ApJS}

\def\gram{\hbox{g}}
\def\cm{\hbox{cm}}

\def\AU{\hbox{AU}}

\def\in{\mathrm{in}}
\def\out{\mathrm{out}}

\def\rad{\mathrm{rad}}
\def\vrt{\mathrm{vert}}
\def\vert{\mathrm{vert}}
\def\modelflare{1}
\def\modelshade{2}
\def\modelthin{3}
\def\modeloutershade{4}
\def\comma{\,,}
\def\fullstop{\,.}
\begin{document}
\title{The 2-D structure of dusty disks around Herbig Ae/Be stars}
\subtitle{I. Models with grey opacities}
\authorrunning{Dullemond}
\author{C.P.~Dullemond}
\titlerunning{The 2-D structure of dusty disks around Herbig Ae/Be stars}
\institute{Max Planck Institut f\"ur Astrophysik, P.O.~Box 1317, D--85741 
Garching, Germany; e--mail: dullemon@mpa-garching.mpg.de}
\date{DRAFT, \today}
\abstract{
In this paper the two-dimensional structure of protoplanetary disks around
Herbig Ae/Be stars is studied. This is done by constructing a
self-consistent model based on 2-D radiative transfer coupled to the
equation of vertical hydrostatics. As a simplifying assumption a grey
opacity is used. It is found that the disk can adopt four different
structures, dependent on the surface density distribution $\Sigma(R)$ as a
function of radius, i.e.~on radial- and vertical optical depth of the
disk. For the case of high to intermediate vertical optical depth, the
temperature and density structures are in agreement with the simple ``disk
with inner hole'' model of Dullemond, Dominik \& Natta (2001, henceforth
DDN01). At large radii the disk adopts a flaring shape as expected, and near
the dust destruction radius (located at about $0.5\AU$ for most Herbig Ae
stars) the disk is superheated and puffs up. The region directly behind this
``puffed-up inner dust wall'' is shadowed, as predicted by DDN01. For the
case of intermediate to low vertical optical depth, but still high radial
optical depth, the 2-D models show that the shadow can cover the entire
disk. For such competely self-shadowed disks the inner rim emission in the
near infrared constitutes the dominant part of the SED, since the flaring
component in the mid- and far infrared is suppressed by the self-shadowing
effect. When the disk is optically thin even in radial direction, it becomes
unshadowed again because the inner rim can no longer block the stellar
light. Such disks have relatively weak infrared excess compared to the
stellar flux. Finally, for disks that flare at intermediate radii, but
become too optically thin at large radii, the outer parts once again become
shadowed. But this time the shadowing is caused by the flaring part of the
disk, instead of the inner rim. The disk then consists of a bright inner rim,
a shadow, a flaring part and finally a (dim) shadowed outer part. Different
observational methods of determining the size of the disk (e.g.~from the SED,
from continuum mapping or from CO mapping) may yield different results.
}

\maketitle


\section{Introduction}
Herbig Ae/Be stars are widely regarded as the intermediate mass counterparts
of the lower mass T Tauri stars (e.g.~Strom et
al.~\citeyear{stromstrom:1972}; Palla \& Stahler
\citeyear{pallastahler:1993}; v.d.~Ancker et
al.~\citeyear{anckerwinter:1998}). But, while there is little doubt that the
infrared excess of classical T Tauri stars originates in a protoplanetary
disk, the case for Herbig Ae/Be stars is not so clear. On large scales the
disk-like nature of the circumstellar matter around Herbig Ae/Be stars is by
now well established. Millimeter CO and continuum maps (Mannings \& Sargent
\citeyear{mannsarg:1997}), submm position-velocity maps (Qi
\citeyear{qithesis:2001}) and imaging at visible and near-infrared
wavelengths (Grady et al.~\citeyear{gradywoodbruh:1999}; Augereau
\citeyear{augereau:2001}) clearly show rotating flattened structures at
scales of hundreds of AU. But on scales of tens of AU and smaller there
seems to be conflicting evidence, with some authors arguing for a more or
less spherical geometry (di Francesco et
al.~\citeyear{difrancescoevans:1994}; Pezzuto et
al.~\citeyear{pezzutostraf:1997}; Miroshnichenko et
al.~\citeyear{miroiveeli:1997},\citeyear{miroivevinkeli:1999}; Millan-Gabet
et al.~\citeyear{millanschl:2001}) and others favoring the disk-like picture
(Vink et al.~\citeyear{vinkdrew:2003}; Grinin \& Rostopchina
\citeyear{grininrostop:1996}; Natta et al.~\citeyear{nattaprusti:2001};
Tuthill et al.~\citeyear{tutmondan:2001}).

Unfortunately, for some time disk models have failed to explain the SEDs of
Herbig Ae/Be stars, and were consequently rejected by many. In particular
the conspicuous bump around 3 microns in the SEDs of Herbig Ae/Be stars
remained a mystery. Standard disk models such as those of Chiang \&
Goldreich (\citeyear{chianggold:1997}, henceforth CG97), and D'Alessio et
al.~(\citeyear{dalessiocanto:1998},\citeyear{dalessiocalvet:1999}) failed to
explain this striking feature. Several explanations were suggested, ranging
from FeO dust at 800 K (v.d.~Ancker et al.~\citeyear{vdanckerbouw:2000}) to
accretion disks that are actively dissipating only beyond a certain radius
(Hillenbrand et al.~\citeyear{hillenstrom:1992}). None of these explanations
were quite satisfactory. Recently, it was recognized by Natta et
al.~(\citeyear{nattaprusti:2001}) that this 3 micron bump may well originate
from the inner rim of the dusty part of the disk. Since dust evaporates
above about 1500 K, the inner parts of the disk are free of dust. These
gaseous inner parts have a much lower optical depth, and can even be
entirely optically thin, dependent on the gas surface density. At the dust
evaporation radius the dust forms a wall of 1500 K that is directly
irradiated by the central star. This produces an extra component in the
spectrum that was not considered by the existing models of flaring disks.

Dullemond, Dominik \& Natta (\citeyear{duldomnat:2001}, henceforth DDN01)
adapted the CG97 model to self-consistently include this inner rim,
and showed that the SEDs of Herbig Ae/Be stars can be naturally explained in
this way. In a recent paper (Dominik et al.~\citeyear{domdulwatwal:2002})
the sample of 14 Herbig Ae/Be stars of Meeus et
al.~(\citeyear{meeuswatersbouw:2001}) was analyzed in the context of this
model, and it was found that the SEDs of most stars were indeed consistent
with the DDN01 picture.

According to this model, the inner rim of the flaring disk is much hotter
than an ordinary flaring disk model would predict at that radius.  This is
because the inner rim is irradiated frontally rather than at a grazing
angle. As a consequence, the inner rim is puffed up and casts a shadow over
part of the flaring disk behind it. This shadow can extend from the inner
rim, at about 0.5 AU, out to 5 AU or more. Outward of this shadowing radius
the disk adopts the usual flaring shape as described by CG97. This part
of the disk is responsible for the observed emission at long wavelengths.
Dependent on the height of the inner rim, the shadow can reach so far out
that the 10 $\mu$m silicate emission feature, produced by warm dust in the
surface layers, is suppressed. This has been used by DDN01 as a possible
explanation for the lack of 10 micron feature in several sources.

Though succesful in explaining several features of the SEDs of Herbig Ae/Be
stars, the DDN01 model was based on highly simplified equations. Among other
things the structure of the inner rim and the shadowing of the disk behind
it need closer theoretical examination. It is unclear from the DDN01 model
what happens when the optical depth of the disk becomes too low to sustain
flaring. Also, the DDN01 model was based on the assumption that, if the disk
{\em can} flare outside the shadow, then it {\em will}. It is unclear
whether perhaps in addition to these flaring disks also fully self-shadowed
disk solutions exist.

Because of the intrinsic 2-D nature of the problem, a closer theoretical
study requires a full 2-D treatment of radiative transfer. This is done with
a 2-D ``Variable Eddington Tensor'' solver. By coupling the radiative
transfer to the equations of vertical hydrostatic equilibrium, the code
solves the entire temperature and density structure of the disk as a
function of radius and vertical height above the midplane. As a simplifying
assumption a grey opacity is adopted in this work. This is consistent with
the disk consisting of large grains. The advantage of this simplification is
that the results are more readily understood in terms of simple radiative
transfer arguments. In a follow-up paper more realistic opacities and grain
size distributions will be used, which will put us in a position to compare
the results directly to observations.

Using the 2-D disk structure code just described, the following issues
will be addressed:
\begin{enumerate}
\item Does the overall structure and SED of the disk agree with the much
simpler model of DDN01? Will there indeed be a shadowed region behind the
inner rim, as predicted by DDN01? How much will radial radiative diffusion
heat this shadowed region, and how large is this region in radial
direction?
\item Under which conditions will the outer part of the disk collapse into
the shadow of the inner rim, making the disk entirely self-shadowed? Can
there be a bimodal set of solutions (self-shadowed and flared) for the same
parameters?
\item What is the structure of a fully self-shadowed disk? Will it be very
cool and collapsed, or does radial radiative diffusion keep the
disk still relatively warm?
\item What will the outer (tenuous) part of a large disk look like: will it
continue to be flaring, or will it sink into the shadow of the flaring part
of the disk?
\end{enumerate}

The paper is organized as follows. In Sec.~\ref{sec-model-equations} the
equations are presented, and it is described how they are solved. In
Sec.~\ref{sec-model-haedisk} a model of a canonnical flaring disk around a
Herbig Ae star is described, and in Sec.~\ref{sec-lowtau-selfshad} it is
shown that self-shadowed disk can exist and what their structure looks
like. In Sec.~\ref{sec-bimodal} it is investigated if multiple solutions
exist for the same parameters.  A description of the tenuous outer parts of
a flaring disk is given in Sec.~\ref{sec-outer-parts}.

\section{The model equations}
\label{sec-model-equations}
The model is based on 2-D radiative transfer to calculate the dust (and
gas) temperature in the disk. Once the temperature is found it is then used
to integrate the equations of hydrostatic equilibrium in order to find the
density structure. The entire procedure is then iterated to obtain the full
disk structure. Since the accretion rate in these disks is presumably very
low, we do not include viscous dissipation in the equations.

The model equations and the computational method used in this work are quite
similar to the ones used for the 1+1-D (vertical) structure models presented
in Dullemond, van Zadelhoff \& Natta (\citeyear{dulvzadnat:2002}). However,
here this method is extended to full 2-D. Details of this method will be
presented elsewhere (Dullemond in prep.).

\subsection{2-D Radiative transfer}
Consider a gas + dust density distribution $\rho(R,\Theta)$ as a function of
the spherical coordinates $R$ (radius) and $\Theta$ (angle as measured from
the polar axis). A grey opacity $\kappa_\nu=\kappa$ is adopted, and the
scattering albedo is assumed to be zero. The star is assumed to be a point
source. The flux of direct starlight $F^{*}(R,\Theta)$ at every point in the
disk is then given by:
\begin{equation}
F^{*}(R,\Theta) = \frac{L^{*}}{4\pi R^2} \exp(-\tau(R,\Theta))
\comma
\end{equation}
where $\tau(R,\Theta)$ is the optical depth in the radial direction between the
point $(R,\Theta)$ and the central star, obeying the conditions:
\begin{equation}
\frac{\partial\tau(R,\Theta)}{\partial R} = \rho(R,\Theta)\kappa
\comma
\end{equation} 
and $\tau(0,\Theta)=0$. This direct stellar radiation is absorbed by the
disk. The amount of energy per unit time and volume that is absorbed in this
way is:
\begin{equation}
Q(R,\Theta) = \rho(R,\Theta)\kappa F^{*}(R,\Theta)
\fullstop
\end{equation}
By virtue of energy conservation this energy must then be reemitted as
thermal (infrared) radiation by the dust in the disk. Dependent on the
optical depth of the disk, this infrared radiation may be absorbed and
re-emitted by the disk many times more, before it eventually escapes to
infinity. It is this radiative diffusion process that allows the radiation
to enter deep into the disk, and determine the temperature at every
location, no matter at which optical depth. 

To solve this complex 2-D radiative transfer problem, the re-emitted
radiation field is treated as a separate radiation field, which will be
called the ``reprocessed radiation field'' from now on. The dust grains
acquire a temperature such that they emit exactly the same energy per second
as they absorb:
\begin{equation}\label{eq-therm-balance}
\frac{\sigma}{\pi} \rho \kappa T^4 = \frac{Q}{4\pi} + \rho\kappa J
\comma
\end{equation}
where $J(R,\Theta)$ is the frequency-integrated mean intensity of the
reprocessed radiation field. The mean intensity $J(R,\Theta)$ is defined as
\begin{equation}\label{eq-meanint}
J = \frac{1}{4\pi}\int_{4\pi} I(\Omega) d\Omega
\comma
\end{equation}
where $I(\Omega)$ is the frequency-integrated intensity of the reprocessed
radiation in the direction $\Omega$. The first term in
Eq.(\ref{eq-therm-balance}) accounts for the absorbed direct stellar
radiation, while the second term takes into account the infrared radiation
emitted by the disk itself. 

The radiative transfer equation for the reprocessed radiation is:
\begin{equation}\label{eq-formal-trans-eq}
\frac{dI(\Omega)}{ds} = \frac{\sigma}{\pi}\rho\kappa T^4 - \rho\kappa I(\Omega)
\comma
\end{equation}
which must be satisfied along every ray through the medium. Here $s$ is the
path length along the ray. The full 2-D radiative transfer problem amounts
to solving the coupled set of equations (\ref{eq-therm-balance},
\ref{eq-meanint}, \ref{eq-formal-trans-eq}). The formal transfer equation
(Eq.~\ref{eq-formal-trans-eq}) has to be integrated along all rays through
the disk. In practice this means that a discrete (but large) set of rays is
chosen which samples the reprocessed radiation field as accurately as
possible. The integration of Eq.(\ref{eq-formal-trans-eq}) along all these
rays is done using the method of ``Extended Short Characteristics''
(Dullemond \& Turolla \citeyear{dultur:2000}).

Solving this set of equations is normally done using an iterative procedure
from Eq.(\ref{eq-formal-trans-eq}) to Eq.(\ref{eq-meanint}), to
Eq.(\ref{eq-therm-balance}), and back to Eq.(\ref{eq-formal-trans-eq}),
until convergence is reached. This procedure is known as ``Lambda
Iteration'' and is the basis of most current 2-D radiative transfer codes
(including the original version of the program {\tt RADICAL} described by
Dullemond \& Turolla \citeyear{dultur:2000}). However, at large optical
depth this method leads to convergence problems since information about the
radiation field propagates only one mean free path per iteration. This can
be fatal for problems involving protoplanetary/protostellar disks. A
rigorous solution to this problem is the ``Variable Eddington Tensor''
method, which couples the formal radiative transfer equation to the
frequency-integrated moment equations (see e.g.~Mihalas \& Mihalas
\citeyear{mihalmihal:1984}; Malbet \& Bertout \citeyear{malbetbertout:1991};
Stone, Mihalas \& Norman \citeyear{stonemihnor:1992}; Dullemond, van
Zadelhoff \& Natta \citeyear{dulvzadnat:2002}). {\tt RADICAL} is now
equipped with this method and can be used to solve 2-D radiative transfer
problems at arbitrary optical depth with only a few iterations\footnote{See
web page {\tt http://www.mpa-garching.mpg.de/
PUBLICATIONS/DATA/radtrans/radical/}.}.

\subsection{Vertical pressure balance}
Once the temperature structure of the disk is determined, one can recompute
the density structure by integrating the equations of vertical hydrostatic
equilibrium:
\begin{equation}\label{eq-vert-hydrostat}
\frac{\partial P}{\partial z}= -\rho\frac{GM_{*}}{R^3} z
\comma
\end{equation}
where $z$ is defined here as $z=R(\pi/2-\Theta)$.  Geometric correction
factors are neglected in the above equation, i.e.~it is assumed that the
disk is still thin enough that these factors are unneccessary.

In order be able to solve Eq.(\ref{eq-vert-hydrostat}), the surface density
$\Sigma(R)=\int_{-\infty}^{\infty}\rho(R,z)dz$ needs to be specified at
every radius. For a given value of $\Sigma(R)$, the vertical density
structure can be found by integrating Eq.~(\ref{eq-vert-hydrostat}) first
for an initial guess of $\rho(z=0)$, and then renormalize it such that the
required surface density is obtained. Having thus found the new density
structure in this way, one can repeat the radiative transfer calculation
for the next iteration step. Only when the new density structure is to
within $10^{-2}$ of the previous density structure, this iteration procedure
is terminated, and a solution is found.

The function $\Sigma(R)$ is an input to the problem, and is entirely free to
choose. This reveils an intrinsic weakness of models of passive disks: since
one is not constrained by a global (constant) accretion rate, one has no
theoretical constraint on this input function. In principle this introduces
an infinite number of degrees of freedom. In practice this problem is less
severe. Disks are presumably formed with a relatively smooth surface density
profile. And the SED of a passive disk depends only weakly on small scale
variations in the surface density. As will be shown below, significant
changes occur only if large parts of the disk switch from optically thick to
optically thin or vice versa.

\subsection{Determining the SED}
Although the simplifying assumption of grey opacities is used in this paper,
it is still useful to calculate the SED from the 2-D disk models. This
is done by making images of the disk at a number of frequencies, and
integrating these images to obtain the fluxes. The images are made using the
ray-tracing capabilities of {\tt RADICAL}. Care is taken that all spatial
scales are sufficiently finely sampled.

Computing the SED at a large number of inclination angles allows for a very
powerful self-consistency check of the radiative transfer solution: the
total outcoming luminosity of the system should be equal to the luminosity
of the star. At the end of each run {\tt RADICAL} carries out such an energy
conservation check. For all computed models, errors remained within 5\%
relative to the infrared luminosity of the disk.

\section{A flaring disk around a Herbig Ae star}
\label{sec-model-haedisk}

As a first step, the structure of a ``typical'' flaring disk around a Herbig
Ae star is computed. As mentioned above, the surface density function
$\Sigma(R)$ is a parameter of the model. It is given as a power law of the
type:
\begin{equation}
\Sigma(R) = \Sigma_0 \; (R/\AU)^p
\end{equation}
where $p$ is the power law index. Furthermore an inner radius $R_{\in}$ and
an outer radius $R_{\out}$ is chosen, but it is verified that the inner
radius $R_{\in}$ produces a dust temperature that is consistent with the
1500 K dust evaporation temperature for silicates. For most models this is
at $0.7\AU$, but for the optically thin model (model \modelthin{}) a smaller
inner radius is required. The parameters of the models are listed in Table
\ref{tab-models}. The one discussed in this section is model \modelflare{}.

\begin{table*}
\centerline{\begin{tabular}[tb]{c|ccc|ccccc|l}
Model &  $M_{\star}$ & $R_{\star}$ & $T_{\rm eff}$ & 
         $\tau_{\vrt}(50\AU)$ & $\tau_{\rad}$ & $p$ & 
         $R_{\mathrm{in}}$ & $R_{\mathrm{out}}$ & Type \\
\hline
1 & 2.0 & 3.0 & 10000 & $1.9\times 10^2$  & $1\times 10^5$ & -1    & 0.7 & 100 & Flaring           \\
2 & 2.0 & 3.0 & 10000 & $2.2$             & $5\times 10^4$ & -2    & 0.7 & 100 & Self-shadowed     \\
3 & 2.0 & 3.0 & 10000 & $7\times 10^{-4}$ & $1$            & -1    & 0.3 & 100 & Very tenuous disk \\
4 & 2.0 & 3.0 & 10000 & $9.8$             & $4\times 10^3$ & -1,-2 & 0.7 & 1000 & Flaring + shadowed \\
\end{tabular}}
\caption{The parameters for the models in this paper. Column 2,3,4 are the
stellar parameters: mass in units of $M_{\odot}$, radius in units of
$R_{\odot}$ and effective temperature in Kelvin. Column 5 is the
vertical optical depth from $\Theta=0$ to $\Theta=\pi$ at a radius of 1
AU. Column 6 is the powerlaw index for the surface density such that
$\tau(R)=\tau(1\AU)*(R/\AU)^p$. Column 7,8 are the inner and outer radius of
the disk in $\AU$. Colum 9 is for comments.}
\label{tab-models}
\end{table*}

\begin{figure}
\centerline{
\includegraphics[width=9cm]{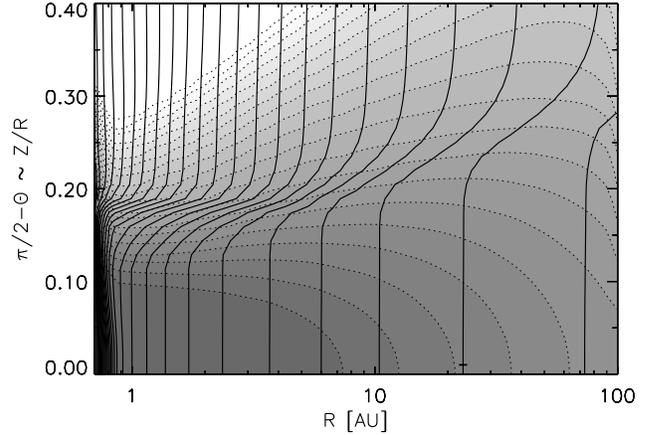}}
\caption{The 2-D temperature and density structure of a flaring disk: model
\modelflare{} discussed in Sec.\ref{sec-model-haedisk}. Contour lines
represent the temperature and are spaced 50 K apart. A small '-'
symbol marks the 100 K contour. Grey scales are density contours, logarithmically
spaced such that 2 grey scale steps represent a factor of 10 in
density. Dotted contours follow the grey scale contours, to aid the eye. The
density contours are stopped at a gas density $\rho=10^{-8}\gram/\cm^3$ to
avoid color crowding. The horizontal axis is the (spherical) radius, in
logarithmic scale. The vertical axis is the vertical height above the
midplane measured in the angle $\pi/2-\Theta$, which is (to first
approximation) equal to the vertical height $z$ divided by the radius
$R$. The equatorial plane is at the lower boundary of the figure. Note that
in the coordinate system used here the stellar (primary) radiation moves
along horizontal lines from left to right (i.e.~at constant $\Theta$).
Therefore, the shadow cast by the inner rim can be recognized as a steep
jump in the temperature along an exactly horizontal line in the figure (at
$\pi/2-\Theta=0.18$, ranging from the inner edge up to 6 AU).}
\label{fig-2d-temp-haedisk}
\end{figure}

In Fig.~\ref{fig-2d-temp-haedisk} the temperature and density structure of
the disk is shown. The disk is considerably cooler at the midplane than at
its surface. This is because the matter deep within the disk is shielded
from direct stellar light, and is only heated indirectly by the reprocessed
radiation field. The inner rim, on the other hand, is directly exposed to
the stellar radiation field, and is therefore much hotter than the rest of
the disk. Directly behind the inner rim the temperature drops strongly, down
to values much smaller than the optically thin dust temperature at the same
radius. 

Between 6 AU and the outer edge the disk adopts a typical flaring shape. The
flaring of the disk can be recognized by following the kink in the
temperature gradient. The relative height ($z/R$) of this kink increases
with radius, showing that the surface height of the disk increases faster
than $R$. This is the typical signature of flaring. This smooth temperature
kink is the ``surface layer'' in which direct stellar radiation is absorbed
and re-emitted as infrared radiation. Rather than a constant temperature (as
in the simple model of Chiang \& Goldreich and also DDN01), this surface
layer has a temperature gradient going from the high temperatures of
grains above the disk down to the lower temperatures of the grains deep
within the disk.

The structure of the disk between the inner edge and about 6 AU is different
from the usual flaring geometry. In the figure one sees that the vertical
temperature gradient suddenly acquires a very steep jump at about $z/R\simeq
\pi/2-\Theta=0.18$. This is a result of the shadow of the inner rim. Below
this line there is no direct stellar radiation, while above it a dust grain
would be exposed to the full stellar flux. In the outer regions (beyond 6
AU) the surface of the disk lies above this shadow line, and therefore
acquires the usual flaring disk shape. But at radii smaller than 6 AU the
disk lies fully in the shadow, and is heated only by radial radiative
diffusion. Note that, in the coordinate system used in
Fig.~\ref{fig-2d-temp-haedisk}, stellar radiation moves along horizontal
lines from left to right.

The reason why the inner rim casts a shadow over the disk behind it is that
the inner rim, being directly exposed to the stellar flux, is much hotter
than the disk behind it, and consequently has a higher pressure scale
height. This hot inner rim is therefore puffed up.

\begin{figure}
\centerline{
\includegraphics[width=9cm]{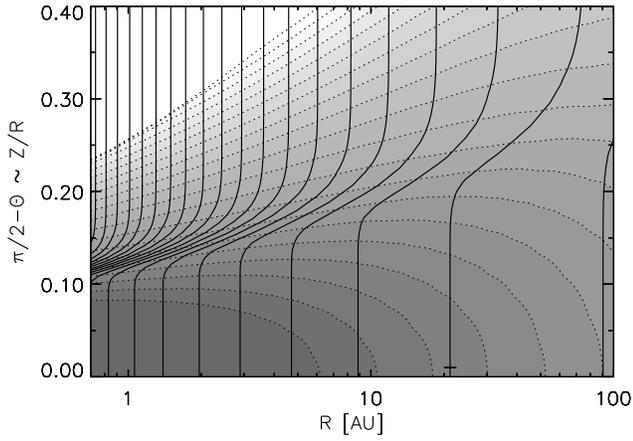}}
\caption{Same as Fig.~\ref{fig-2d-temp-haedisk}, but now computed using
the standard 1+1-D approach to radiative transfer. It is clear that the
effects of the hot inner rim and the shadowed region are not present 
using this 1+1-D approach.}
\label{fig-1-1d-temp-haedisk}
\end{figure}

In Fig.~\ref{fig-1-1d-temp-haedisk} the same disk model was computed using
the standard 1+1-D splitting, i.e. vertical plane-parallel radiative transfer
(see e.g.~Malbet et al.~\citeyear{malbetbertout:1991}; Dullemond, van
Zadelhoff \& Natta \citeyear{dulvzadnat:2002}). This model nicely reproduces
the flaring shape of the disk. But it fails to reproduce the heating of the
inner rim and the shadowed region. This shows that, while 1+1-D models can
be used reasonably well to model disks at radii larger than several AU, 
they cannot be used for the inner regions.

\begin{figure}
\centerline{
\includegraphics[width=9cm]{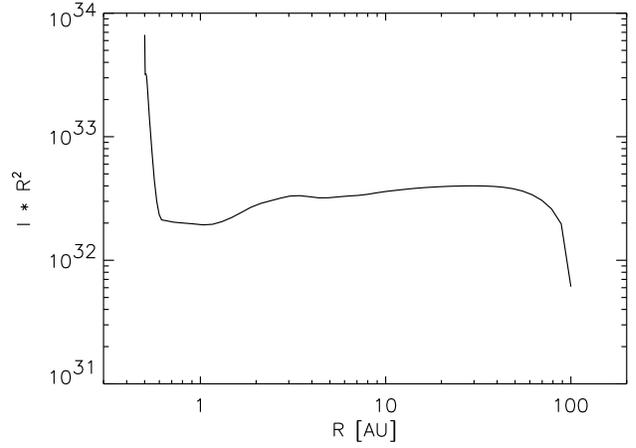}}
\caption{The frequency-integrated intensity of the flaring disk (model
\modelflare{}) seen face-on, as a function of radius. This is computed by
integrating the formal transfer integral (Eq.~\ref{eq-formal-trans-eq})
vertically through the disk. The multiplied by $R^2$ in order to weigh it
with the emitting surface, so that it represents how strong the emission at
each radius contributes to the SED. It is, so to say, the emission of the
disk per constant interval in $\log(R)$.}
\label{fig-emis-shadow}
\end{figure}

Though the shadow is clearly seen in Fig.~\ref{fig-2d-temp-haedisk}, in the
shadowed region the midplane temperature does not drop very much.  In fact,
the temperature still increases as one goes towards smaller $R$.  This shows
that radial radiative diffusion manages to smear out radiative energy quite
efficiently, so that even matter in the shadow of the inner rim (all the
matter with $\pi/2-\Theta\lesssim 0.17$ in this case) will still be
relatively warm, though colder than what it could be if the surface of the
disk were out of the shadow. The thermal emission from this shadowed region
is somewhat suppressed, as can be seen in Fig.~\ref{fig-emis-shadow}.  This
suppression is less than what was expected by DDN01.  This is
because DDN01 used a very approximate formula to estimate the effect of
radial radiative diffusion. In the present 2-D calculations the effect of
this radial diffusion turns out to be stronger than predicted by DDN01.

From Fig.~\ref{fig-emis-shadow} it is also interesting to note that the
emission per unit of $\log(R)$ is more or less constant for the flaring part
of the disk (from about 3 AU onwards, up to shortly before the outer edge at
100 AU). Every radius of the flaring disk constributes equally to the
SED. This is a feature that can be understood in terms of geometry. The
surface height $H_s$ of the disk goes roughly as $H_s\propto R^{9/7}$ (see
CG97). This means that the flaring surface of the disk captures stellar
luminosity in a way that is proportional to $R^{2/7}$. This is only a slowly
varying function of $R$. In Secs.~\ref{sec-lowtau-selfshad} and
\ref{sec-outer-parts} two cases will be discussed where the flaring geometry
of the disk breaks down. And indeed, it will turn out that the ``equal
emission per $\log(R)$'' will also cease to hold. \revised{Note
that at the very outer edge of the flaring disk the emission drops somewhat
(Fig.~\ref{fig-emis-shadow}). This is a result of radiative leaking of 
energy through the outer edge, and is therefore a 2-D radiative transfer
effect.}

\begin{figure}
\centerline{
\includegraphics[width=9cm]{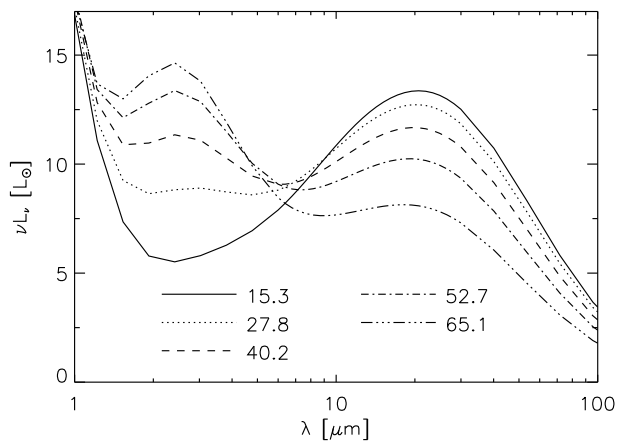}}
\caption{The spectral energy distribution (for grey opacities) for model
\modelflare{}, computed at different inclination angles. The stellar
atmosphere is included, but is approximated by a blackbody for
simplicity. The inclination angles are measure from the pole (i.e.~$i=0$
means face-on). Note that the vertical scale is linear.}
\label{fig-seds-flaredisk}
\end{figure}

In Fig.~\ref{fig-seds-flaredisk} the SED of the 2-D model is shown at
various inclination angles. One can see that for face-on inclinations the
3-micron bump emission from the inner rim is very weak compared to the
emission from the flaring part of the disk at longer wavelengths. This is
because the inner rim is seen along the rim-surface, instead of
perpendicular to it. At larger inclinations the inner rim emission gains in
strength, while the flaring part fades. These are geometrical effects were
predicted by DDN01 and seem to be confirmed here. It should be noted,
however, that the strong supression of the near infrared bump at face-on
inclinations is a result of the perfectly vertical inner rim that is assumed
in our model. In reality, hydrodynamic effects will presumably make this
inner rim more round (as in Fig.~\ref{fig-pictograms} below), which allows
the near infrared emission to be seen also at face-on inclinations. As there
is observational evidence that even face-on disk show considerable
near-infrared emission (e.g.~Millan-Gabet et
al.~\citeyear{millanschl:1999}), it is important to investigate such
hydrodynamic ``smoothing'' of the inner rim in more detail in the near
future.

\begin{figure}
\centerline{
\includegraphics[width=4cm]{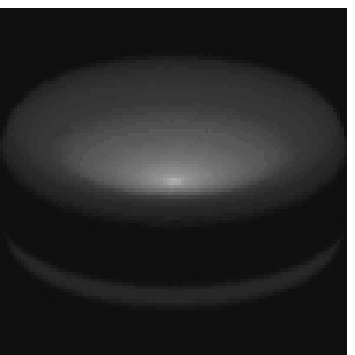}
\includegraphics[width=4cm]{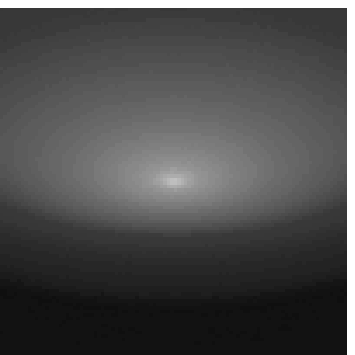}}
\vspace{0.2em}
\centerline{
\includegraphics[width=4cm]{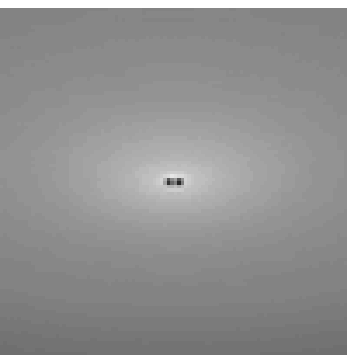}
\includegraphics[width=4cm]{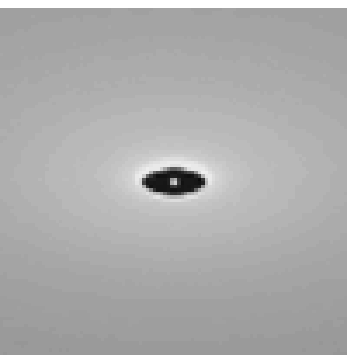}}
\vspace{0.2em}
\centerline{
\includegraphics[width=4cm]{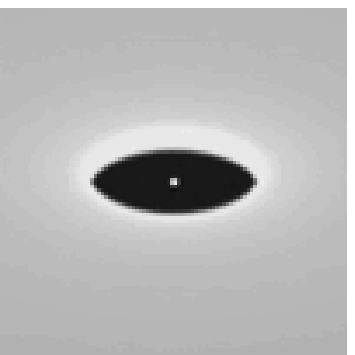}
\includegraphics[width=4cm]{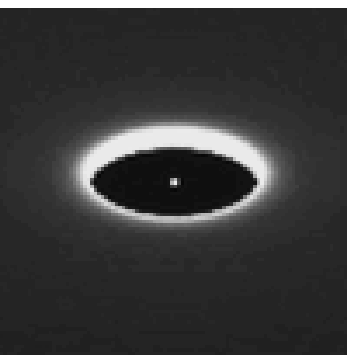}}
\caption{Images of the flaring disk (model \modelflare{}) at varying zoom
factor. All images, with the exception of the bottom-right image, are in
logarithmic grey scale. The bottom-right image is the same as the bottom-left
image, but then in linear grey scale. From top-left to bottom-right the
images show the entire disk, zoomed in to the bright inner rim.
The images are taken at 50 $\mu$m.}
\label{fig-model-flare-images}
\end{figure}

Images of how the disk actually looks like when viewed at a certain
inclination angle are shown in Fig.~\ref{fig-model-flare-images} in
logarithmic grey scale. A wavelength of 50 $\mu$m is chosen so that the
entire disk radiates sufficiently brightly. In the upper-left image one sees
the outer edge of the disk. It is evident that the surface layers radiate
strongly, while the midplane layers are dimmer. As one zooms in, the
intensity of the disk's surface gets stronger, and finally the bright inner
rim appears. The inner rim is much brighter than the rest of the disk, as
can be seen in linear grey scale (the lower-right image). The far side of
the rim provides most of the flux.  The inner rim therefore appears as an
ellipse with one side brighter than the other. This asymmetry will
presumably be less strong if one takes into account the smoothing of the
inner rim by hydrodynamic effects, as discussed above. A strong shadowed
region does not show up clearly in these images, dispite the fact that the
shadow is indeed there (see Figs.~\ref{fig-2d-temp-haedisk} and
\ref{fig-emis-shadow}).  Clearly the radial radiative diffusion prevents
the shadowed region from becoming too cold.

\section{Self-shadowed disks}
\label{sec-lowtau-selfshad}

\begin{figure}
\centerline{
\includegraphics[width=9cm]{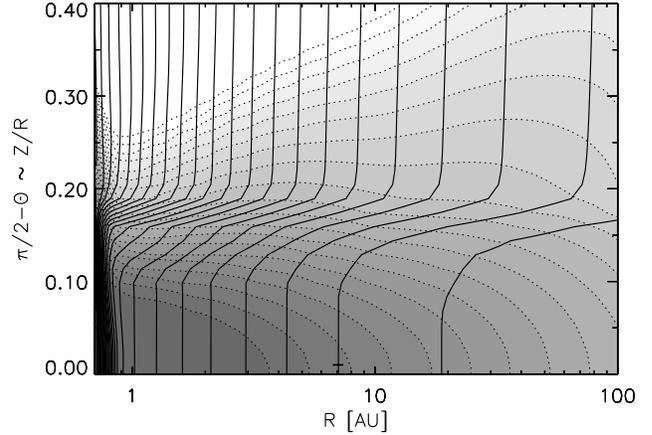}
}
\caption{The temperature and density structure of a self-shadowed disk:
model \modelshade{} discussed in Sec.~\ref{sec-lowtau-selfshad}. Contour
lines represent the temperature and are spaced 50 K apart. A small '-'
symbol marks the 100 K contour. Grey scales are density contours,
logarithmically spaced such that 2 grey scale steps represent a factor of 10
in density.}
\label{fig-model-selfshad-temp}
\end{figure}

\begin{figure}
\centerline{
\includegraphics[width=9cm]{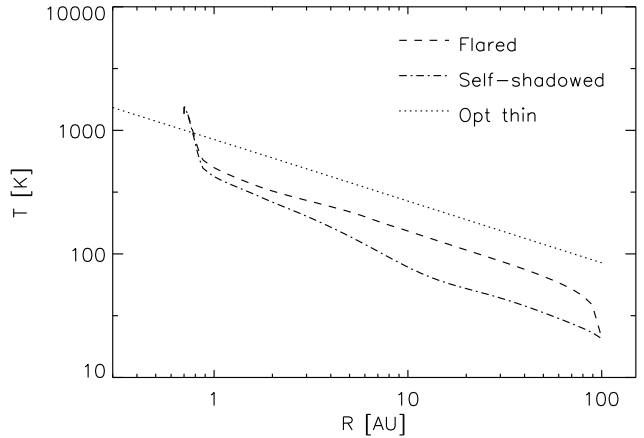}
}
\caption{The temperature at the equator as a function of $R$, for the
self-shadowed disk of model \modelshade{} (\revised{dot-dashed line}) and
for the more massive flaring disk of model \modelflare{} (dashed line).
\revised{The temperature for a completely optically thin configuration is
shown as a dotted line. Note that the temperature of models \modelflare{}
and \modelshade{} near the inner edge (0.7 AU) exceeds the optically thin
dust temperature. This is a backwarming effect by the optically thick inner
rim.}  }
\label{fig-model-selfshad-temp-equator}
\end{figure}

For disks that do not reach a certain minimal mass, it can happen that the
entire disk will sink into the shadow of the inner rim. At no point the
vertical optical depth is then large enough that the surface of the disk
reaches out of the shadow. The disk (apart from the inner rim) is therefore
not directly heated by the stellar radiation, and will therefore cool
down. Since the temperature determines the vertical height, the disk will
therefore shrink even deeper into the shadow. The temperature and density
structure of such a disk is shown in Fig.\ref{fig-model-selfshad-temp}. The
shadow is now clearly seen to extend all the way towards the outer edge of
the disk. Yet, the self-shadowed disk is not entirely cold, as can be more
clearly seen in Fig.~\ref{fig-model-selfshad-temp-equator}. Although no part
of the disk behind the inner rim can see the star directly, there is still
sufficient indirect heating by the reprocessed radiation of the inner parts
of the disk, and by radial radiative diffusion, that the self-shadowed disk
remains relatively warm.

The self shadowed disk has no hot surface layer: the temperature is more or
less the same at every height above the midplane. Only at elevations
significantly above the photosphere of the disk, where one gets in direct
sight of the star, does the temperature rise to the optically thin dust
temperature. But at those elevations the densities are already so low that
one cannot speak of a hot surface layer anymore.

\begin{figure}
\centerline{
\includegraphics[width=9cm]{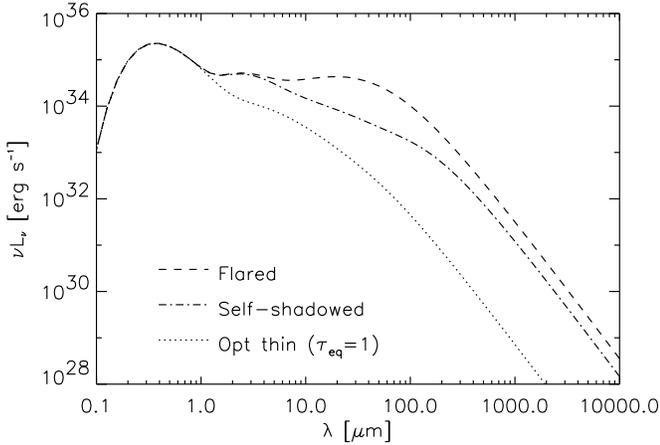}
}
\caption{\revised{The spectral energy distribution of the flaring disk of
model \modelflare{} (dashed line), the self-shadowed disk of model
\modelshade{} (\revised{dot-dashed line}), and the optically thin disk of
model \modelthin{} (dotted line)}.}
\label{fig-sed-shadowed}
\end{figure}

In Fig.~\ref{fig-sed-shadowed} the SED of the self-shadowed disk is shown,
and compared to the SED of a flaring disk. Although this SED has been
computed here only for grey opacities, it is interesting to note that the
SED of the self-shadowed disk has a much steeper slope towards long
wavelengths ($\nu F_\nu\propto \nu^{1.1}$) than the flared disk ($\nu
F_\nu\propto \nu^{0.2}$), which is a result of the steeper temperature slope
of these disks. One may speculate whether such self-shadowed disks may be
the explanation for the group II Herbig Ae/Be stars in the classification of
Meeus et al.(\citeyear{meeuswatersbouw:2001}). These sources show a weak far
infrared emission, and have an overall SED shape very similar to the SED
computed here for the self-shadowed disks. \revised{Moreover, the similar
strength in the near-IR emission between self-shadowed and flared disks is
also observed between group I and group II sources.}  Yet, these group II
sources have strong silicate emission features, which might not be
consistent with such self-shadowed disks. \revised{Also, mm observations
indicate that group II sources are still quite massive, and it remains to be
proven whether these disks are not too massive to be self-shadowed.}  To
answer these questions, a similar disk model, but with more realistic dust
opacities must be carried out. This will be one of the subjects of a
follow-up paper (Dullemond \revised{\& Dominik} in prep.).

\section{Very tenuous disks}
\label{sec-tenuous}

\begin{figure}
\centerline{
\includegraphics[width=9cm]{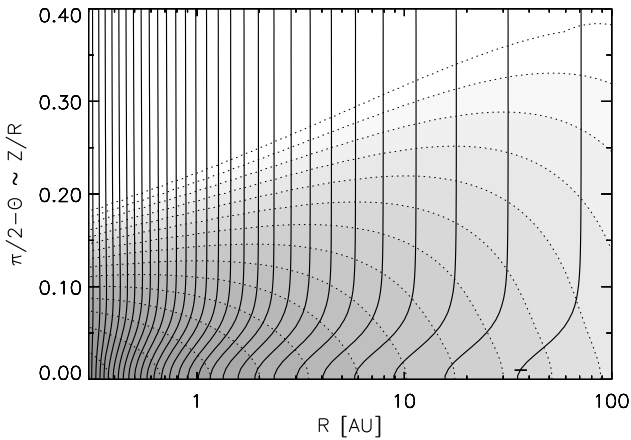}}
\caption{The temperature and density structure of an almost optically thin
disk: model \modelthin{} discussed in Sec.~\ref{sec-tenuous}.  Contour lines
represent the temperature and are spaced 50 K apart. A small '-'
symbol marks the 100 K contour. Grey scales are density contours, logarithmically
spaced such that 2 grey scale steps represent a factor of 10 in density.}
\label{fig-model-thin-temp}
\end{figure}

When the optical depth measured along the equatorial plane becomes smaller
than unity, the height of the shadow drops to zero and the disk becomes
unshadowed again. In this case not only the vertical optical depth
$\tau_\vert$ is smaller than unity, but the optical depth along any ray
through the disk is smaller than one. It can be considered as a fully
optically thin disk. Since such a disk has no surface layer which captures
the stellar radiation (as in model \modelflare{}), the disk is not a
``flaring disk'' as such. Yet, the density structure is still ``flared'' in
another sense: the dimensionless pressure scale height $h_p\equiv H_p/R$
increases outwards, as can be clearly seen from the density contours in
Fig.~\ref{fig-model-thin-temp}. This is easily understood, since the
temperature of a grey dust grain goes as $T\propto 1/\sqrt{R}$ and the
pressure scale height scales as $H_p\propto \sqrt{TR^3}$, giving 
$H_p/R\propto R^{1/4}$. 

Model \modelthin{} is a model with an equatorial optical depth (radial
optical depth along the midplane) of $\tau_\rad=1$. In
Fig.~\ref{fig-model-thin-temp} the temperature and density structure of this
disk is shown. It is clear that the temperature is only modestly influenced
by the extinction at the equator, and that the overall density structure is
that of a purely optically thin disk. The covering fraction of this disk is
very small, and therefore the infrared excess of this disk is small compared
to the stellar flux (Fig.~\ref{fig-sed-shadowed}).

The optically thin model presented here (model \modelthin{}) may not be very
realistic. In such very tenuous disks the temperature of the dust and the
gas may start to decouple. Gas heating/cooling, dust-gas thermal energy
exchange and dust drift should then be properly taken into account (see
e.g.~Kamp \& van Zadelhoff \citeyear{kampzadel:2001}).

\section{Outer regions of flaring disks}
\label{sec-outer-parts}

Generally it is assumed that the surface density $\Sigma(R)$ of a
protoplanetary disk decreases as a function of radius. It can happen that a
disk of the usual ``rim + shadow + flaring'' type will become too optically
thin in the outer parts to maintain flaring beyond a certain radius
$R_{\mathrm{turn}}$. The $\tau_V=1$ surface, where the stellar radiation is
captured by the disk, in effect turns over and no longer maintains the
necessary flaring shape necessary to capture stellar radiation. The hot
surface layer ceases to exist beyond that radius. The outer regions can now
only be heated indirectly by the infrared radiation of the disk itself. This
prevents the disk from collapsing to zero scale height.

Model \modeloutershade{} is a model that has such a non-flaring outer
part. To aggravate the situation so that the results are more clear, the
surface density distribution is taken to be a broken powerlaw ($p=-1$
inwards of 50 AU an $p=-2$ outwards of 50 AU). But the self-shadowed outer
regions can occur also for single-powerlaw surface density slopes, though
the dimming of the emissivity is then weaker.

\begin{figure}
\centerline{
\includegraphics[width=9cm]{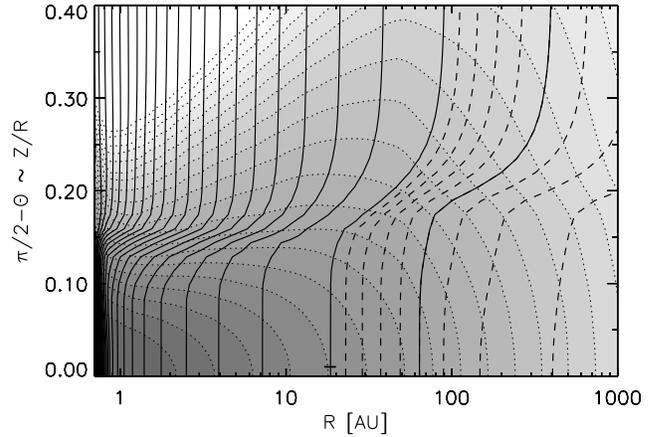}}
\caption{The temperature structure of a large disk which becomes optically
thin to stellar radiation at the outer parts: model \modeloutershade{}
discussed in Sec.~\ref{sec-outer-parts}. Contours are 50 K apart.  A small
'-' symbol marks the 95 K contour. The dashed lines are also temperature
contours, but with a spacing of 5 K, in order to make the structure at large
radii more clear. Note that the radial range is larger than in the previous
figures of this kind.}
\label{fig-model-thinouter-temp}
\end{figure}

\begin{figure}
\centerline{
\includegraphics[width=9cm]{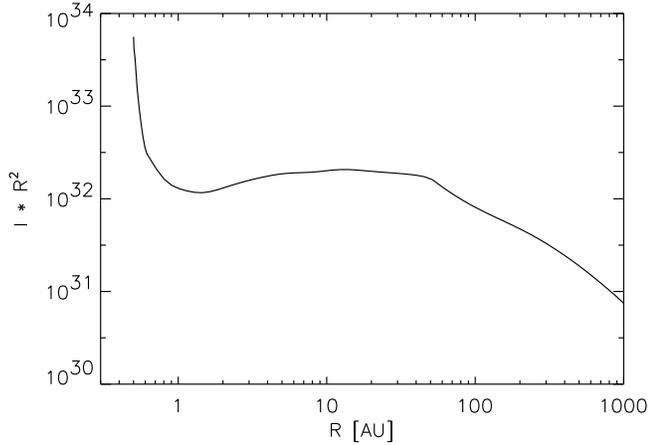}}
\caption{As Fig.~\ref{fig-emis-shadow}, but now for model 
\modeloutershade{}.}
\label{fig-model-thinouter-emis}
\end{figure}

In Fig.~\ref{fig-model-thinouter-temp} the temperature structure of this
disk is shown. The shadowing effect is not very clear from the solid
contours. That is why a number of finer-spaced contours are added in the
figure (dashed lines). The shadow is apparent in the same way as it was for
the shadow of the inner rim: a horizontally arranged temperature jump. But a
better way to see the effect of the shadowing of the outer regions is to
compute the emission of the disk per $\log(R)$, as shown in
Fig.~\ref{fig-model-thinouter-emis}. It is clearly seen that the continuum
emission drops strongly in the shadowed outer regions, and will therefore
only make a small contribution to the SED.

This result may go some way towards explaining the differences that are
often encountered when measuring the radius from SED fitting, mm continuum
interferometric mapping and CO line interferometric mapping (e.g.~Mannings
et al.~\citeyear{mannsarg:1997}; \citeyear{mannkoersarg:1997}; Dominik et
al.~\citeyear{domdulwatwal:2002}). As the results of this model show, all of
the radiation of the star is reprocessed within $R_{\mathrm{turn}}$, since
the disk is incapable of capturing radiation beyond that radius. So the SED
will be dominated by the disk structure within $R_{\mathrm{turn}}$, and the
emission from the outer parts will be weak in comparison. Therefore the
outer radius as inferred from SED fitting will yield
$R_{\mathrm{out}}=R_{\mathrm{turn}}$. But in interferometric maps one can
see also the region beyond $R_{\mathrm{turn}}$, even though it is much
dimmer than the emission from the flaring parts of the disk. CO rotational
lines, since they are density tracers, will be less senstive to this
temperature drop. They may even be more easily detected in these outer
regions, which may yield even larger effective disk sizes, unless CO is
strongly depleted by freezing out onto dust grains.

\section{Bimodal solutions?}
\label{sec-bimodal}

In the paper of DDN01 the possibility was raised that two disk solutions
might exist for the same disk parameters, namely one where the disk has the
rim+shadowed region + flaring part structure, discussed in
Sec.~\ref{sec-model-haedisk}, and one where, on the contrary, the whole disk
is in the shadow of the inner rim, as discussed in
Sec.~\ref{sec-lowtau-selfshad}. This is an important point because, if
proven, it would mean that the actual shape of a disk depends not only on
its properties such as mass and radius, but also on its previous evolution.

The reasoning is as follows. Suppose we do not consider the process of
radial radiative transport, and assume that the interior of the disk is only
heated by radiation from the surface layer that diffuses downwards, but not
sideways to other radial shells. In such a scenario it is unclear why the
disk should always flare (and pop out of the shadow) at radii where it
can. Suppose we could artifically cool the flaring part of the disk down for
a moment. Then the disk will shrink in height and retreat back into the
shadow. Once in the shadow, it won't receive radiation from the star (the
surface remains below the shadow line), and therefore the disk remains cool
and stays in the shadow. Hence we would have found a second solution. In the
DDN01 paper it was argued that presumably nature will pick the flaring
solution only. Yet this remained to be proven.

With the 2-D radiative transfer code one is now in a position to do this,
since 2-D transfer automatically includes the effect of radial radiative
diffusion and indirect heating by radiation from the inner regions. The
parameters of the disk are chosen such that it is only marginally flaring. The
iteration of the disk structure computation was started from two different
initial guesses for the vertical pressure scale height. One was a normal
flaring disk ($H_p\propto R^{2/7}$), and one was a fully self shadowed disk
($H_p\propto R^{-2/7}$).

It was found that both initial guesses reach the same solution, which is a
``rim + shadow + flaring'' disk. This procedure was tried for a series of
parameters and optical depths. In each case only a single solution was
found. This means that the possibility of bimodal solutions seems to be
excluded, at least within the present assumption of grey opacities. The
conclusion is what one could call the ``flaring disk principle'': a disk
will flare whenever and wherever it can.  Only if the disk cannot flare,
because of too low optical depth, then it will remain self-shadowed.

The question of bimodal solutions is also loosely linked to the question of
stability. Based on a time-dependent version of the equations of CG97 for an
irradiated flaring disk it can be shown that such disks should be unstable
to self-shadowing (Dullemond \citeyear{dullemond:2000}; Chiang
\citeyear{chiang:2000}). Such disks would quickly develop ripples on the
surface, which eventually cause most of the disk to collapse. However, these
conclusions were based entirely on the highly simplified equations of
CG97. It appears that 2-D radiative transfer effects will stabilize the disk
(Dullemond in prep.). The radial thermal coupling between neighboring parts
of the disk prevents the development of strong temperature gradients.

\section{Conclusions}
\label{sec-disc-concl}

\begin{figure*}
\mbox{}\vspace{1em}\\
\parbox[t]{8.2cm}{
\centerline{A: Flared disk:}
\centerline{\includegraphics[width=7.5cm]{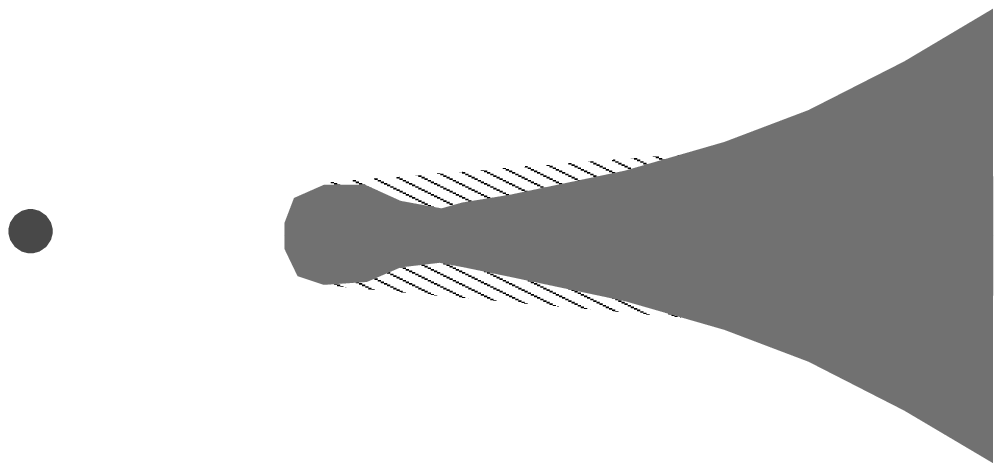}}
}
\parbox[t]{8.2cm}{
\centerline{B: Flared disk with shadowed outer region:}
\centerline{\includegraphics[width=7.5cm]{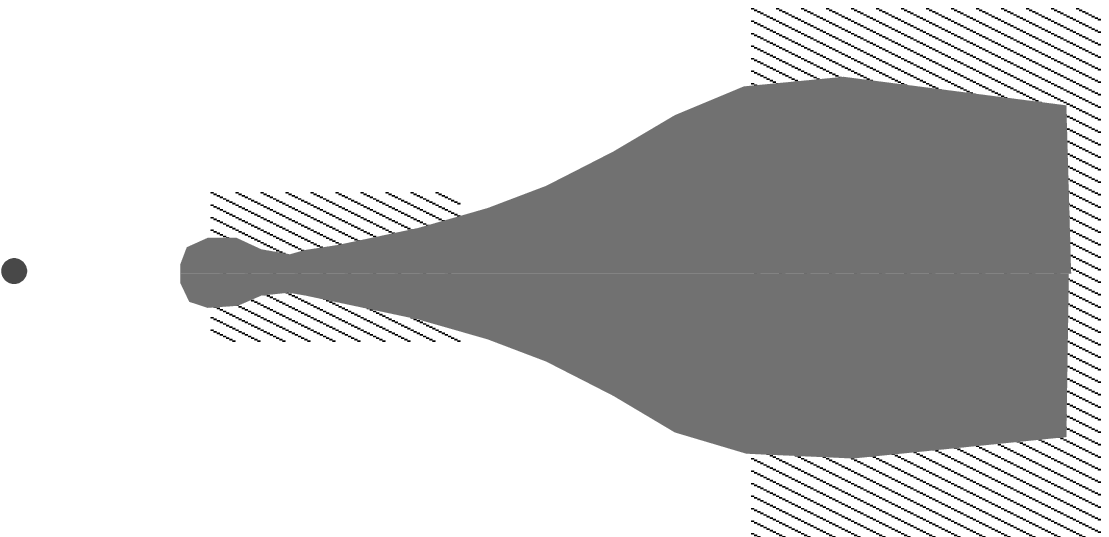}}
}\\
\parbox[t]{8.2cm}{
\centerline{C: Self-shadowed disk:}
\centerline{\parbox[b]{7.5cm}{\mbox{}\vspace{1.0cm}\\
\includegraphics[width=7.5cm]{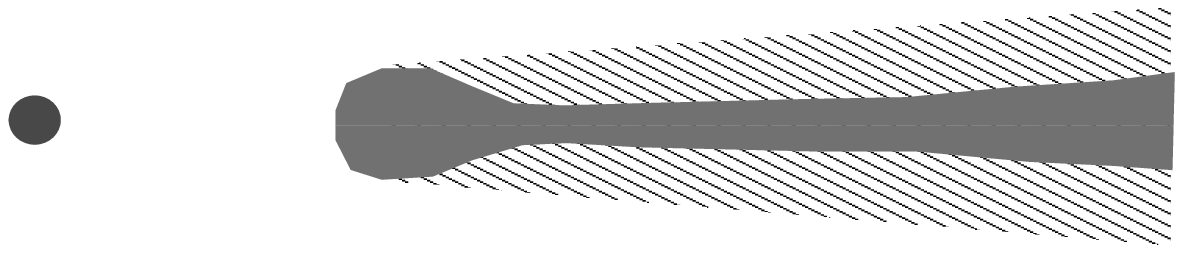}
\\\vspace{1.2cm}\mbox{}}}
}
\parbox[t]{8.2cm}{
\centerline{D: Transparent disk:}
\centerline{\parbox[b]{7.5cm}{\mbox{}\vspace{0.6cm}\\
\includegraphics[width=7.5cm]{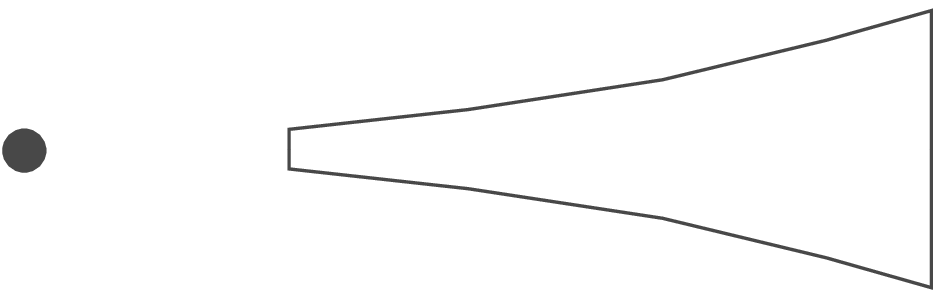}
\\\vspace{0.6cm}\mbox{}}}
}
\caption{Pictograms showing the four main kinds of solutions found.  They
represent a vertical cross-section of the disk, but are not to scale. A
hashed area represents a shadow. In cases A, B and C the vertical optical
depth may, under some conditions, drop below zero even though the radial
optical depth remains much larger than unity. Case B is ``zoomed out'' to
indicate that the second shadowing happens at large radii. In case D, the
empty polygon is meant to show that even the radial optical depth along the
midplane is small, meaning that the temperature everywhere is set by the
optically thin dust temperature. \revised{The cases A, B, C and D are
ordered according to decreasing optical depth (i.e.~mass of the disk). It
should be noted, however, that such an ordering does not always strictly
apply since the powelaw index $p$ of the surface density distribution also
plays a role in determining the disk shape.}}
\label{fig-pictograms}
\end{figure*}

In this paper the structure of passive circumstellar disks was theoretically
investigated. The problem was defined in a mathematically ``clean'' way: it
is the problem of computing the temperature and density structure of
rotating circumstellar matter around a star with a certain mass, radius and
luminosity from basic principles of radiative transfer, radiative
equilibrium and vertical hydrostatic equilibrium. The disk parameters that
went into the calculation were the inner and outer radius, and the surface
density distribution as a function of radius. The only mathematical
approximation made here was the reduction of the hydrostatic equilibrium
equations to 1-D vertical equations. From a physical point of view, many
more approximations were made (related to dust-gas coupling, active
accretion, dust opacities, etc). But these were necessary to keep the
problem clear of uncertain physics for now.

Four different kinds of solutions were found: a flaring disk, a
self-shadowed disk, a transparent disk, and a flaring disk with
self-shadowed outer region. These solutions are pictographically listed in
Fig.~\ref{fig-pictograms}. The numerical models described in this paper can
be downloaded from a website: {\tt
www.mpa-garching.mpg.de/PUBLICATIONS/DATA/ radtrans/grey2d/}.

The main conclusions are summarized as follows:
\begin{enumerate}
\item A flaring disk around a Herbig Ae/Be star has a hot inner rim, a shadowed
region behind it, and the usual flared geometry at large radii
(Fig.~\ref{fig-pictograms}-A). These findings are in accordance with the
predictions of Dullemond, Dominik \& Natta (\citeyear{duldomnat:2001}). But
the effect of shadowing in suppressing the emission from the shadowed region
is not as strong as was predicted in that paper.
\item Disks with intermediate to low vertical optical depth but high
equatorial optical depth can become entirely self-shadowed
(Fig.~\ref{fig-pictograms}-C). The SED falls off more steeply at long
wavelength than for flaring disks.
\item Disks with equatorial optical depths that are smaller than unity
are un-shadowed again, since the inner rim can no longer stop the stellar
radiation. These disks are fully optically thin (Fig.~\ref{fig-pictograms}-D).
\item The outer regions of flared disks can become shadowed if beyond a
certain radius the surface density becomes too low. This time it is the
flared part of the disk that casts the shadow (Fig.~\ref{fig-pictograms}-B).
These outer parts do not contribute much to the SED, but may still be
detectable using (sub-)millimeter interferometers. Measurements of the
outer radius of a disk may therefore yield different results, depending
on whether one uses SED-fitting, continuum mapping or CO mapping.
\item In the case of self-shadowed disks, and the shadowed outer parts of
flaring disks, the usual 1+1-D approach to disk modeling breaks down, and so
does the approach used by CG97 and DDN01. A full 2-D approach, such as the
one used in this paper, is then necessary. It might be that the SED is still
relatively well described using a 1+1-D model or a DDN01-type model up to the
self-shadowing radius, but this remains to be proven using a 2-D model with
more realistic opacities.
\item Bimodel solutions were not found. It seems that protoplanetary disks
obey a kind of ``flaring disk principle'': If the disk {\em can} flare, it
{\em will}. Self-shadowed disks are therefore disks which cannot be made to
flare.
\end{enumerate}

Many of these conclusions will presumably still hold when more realistic
opacities are included. But this will be the topic of the second paper in
this series.

\begin{acknowledgements}
I wish to thank Carsten Dominik and Antonella Natta for their careful
reading of the manuscript and many interesting remarks, Tom Abel for
inspiring me during the debugging of the variable eddington tensor code, and
Rens Waters and G.-J. van Zadelhoff for useful discussions. I acknowledge
support from the European Commission under TMR grant ERBFMRX-CT98-0195
(`Accretion onto black holes, compact objects and prototars').
\end{acknowledgements}

\end{document}